\documentclass[a4paper,11pt]{article}
\usepackage{jcappub}
\usepackage[T1]{fontenc}
\usepackage{amsmath, amssymb, graphics}
\usepackage{mathtools}
\usepackage{tablefootnote}
\usepackage{soul}
\usepackage[normalem]{ulem}
\usepackage{color}
\usepackage{float}
\usepackage{makecell}
\usepackage{booktabs}
\usepackage[usenames,dvipsnames,svgnames,table]{xcolor}
\usepackage{indentfirst}
\usepackage{leftidx}
\usepackage{multirow}
\usepackage{comment}

\definecolor{LightCyan}{rgb}{0.88,1,1}
\newcolumntype{a}{>{\columncolor{LightCyan}}c}

\newcommand{\apjl}{Astrophys. J. Lett.}
\newcommand{\apj}{Astrophys. J.}

\newcommand{\aap}{Astron. Astrophys.}

\newcommand{\mnras}{Mon. Not. Roy. Astron. Soc.}

\newcommand{\jcap}{JCAP}

\usepackage{xspace}

\newcommand{\hmpc}{$h^{-1}\mathrm{Mpc}$\xspace}
\newcommand{\hmsun}{$h^{-1}M_\odot$\xspace}

\title{Neutrino Halo profiles: HR-DEMNUni simulation analysis}

\author[1]{Beatriz Hern\'andez-Molinero,}
\author[2]{Carmelita Carbone,}
\author[3,4]{Raul Jimenez,}
\author[1,5]{Carlos Pe\~na Garay}

\affiliation[1]{Laboratorio Subterr\'aneo de Canfranc, 22880 - Canfranc-Estaci\'on , Huesca, Spain.}
\affiliation[2]{INAF – Istituto di Astrofisica Spaziale e Fisica cosmica di Milano (IASF-MI), Via Alfonso Corti 12, I-20133 Milano, Italy}
\affiliation[3]{ICCUB, University of Barcelona, Marti  i Franques 1, E-08028 Barcelona, Spain.}
\affiliation[4]{Instituci\`o Catalana de Recerca i Estudis Avan\c{c}ats, Pg. Lluis Companys 23, Barcelona, E-08010, Spain.}
\affiliation[5]{I2SysBio, CSIC-University of Valencia, 46071 - Valencia, Spain.}

\emailAdd{bhernandez@lsc-canfranc.es}
\emailAdd{carmelita.carbone@inaf.it}
\emailAdd{raul.jimenez@icc.ub.edu}
\emailAdd{cpenya@lsc-canfranc.es}

\abstract{Using the high-resolution HR-DEMNUni simulations, we computed neutrino profiles within virialized dark matter haloes. These new high-resolution simulations allowed us to revisit fitting formulas proposed in the literature and provided updated fitting parameters that extend to less massive haloes and lower neutrino masses than previously in the literature, in accordance with new cosmological limits. The trend we observe for low neutrino masses is that, for dark matter halo masses below $\sim 4\times10^{14}$\hmsun, the presence of the core becomes weaker and the profile over the whole radius is closer to a simple power law. We also characterized the neutrino density profile dependence on the solid angle within clustered structures: a forward-backward asymmetry larger than 10\% was found when comparing the density profiles from neutrinos along the direction of motion of cold dark matter particles within the same halo. In addition, we looked for neutrino wakes around halo centres produced by the peculiar motion of the halo itself. Our results suggest that the wakes effect is observable in haloes with masses greater than $3\times10^{14}$\hmsun where a mean displacement of $0.06$\hmpc was found.}

\begin{document}
\maketitle	

\section{Introduction}

Cosmology remains the most promising route to prove the absolute mass of neutrinos. Current cosmological surveys are already providing very stringent upper limits that are very close to the total mass inferred from underground experiments that measure the mass splits, i,e, $0.059$ eV~\cite{DESI},  
which in turn confidently suggests, within the framework of Bayesian evidence, that the hierarchy is normal~\cite{Fer1,Fer2}. However, in order to confirm that we have actually discovered neutrinos and not something else, it is important to have additional tests about the nature of this `hot' component of the Universe density budget. Some test can also be obtained by measuring the sound speed and viscosity of this extra component~\cite{licia}. Another important probe is to study the clustering of neutrino haloes around their corresponding cold dark matter ones. This is the probe we explore in this work.

If massive neutrinos exist in the Lambda Cold Dark Matter (LCDM) model, then there should be clear signatures of how they cluster around Cold Dark Matter (CDM) haloes. It is these signatures, which could be measured by e.g. weak lensing surveys, that could provide evidence that we have indeed detected massive neutrinos. In that sense, in the present work we aim to provide high fidelity signatures about neutrino haloes by cold dark matter and neutrino density profiles inside and around virialized haloes, making emphasis in their angular dependence within the neutrino halo. 

In particular, we revise the fitting formulas proposed in~\cite{Villaescusa-Navarro_2013} but using improved simulations which have a significantly larger mass resolution for both Cold Dark Matter (CDM) and neutrino particles, in agreement with current cosmological simulations, which allows us to get more reliable results. Firstly, we wanted to see how previous formulas stand for new simulations where the neutrino mass is twice smaller. We will see how these new simulations push higher the applicability limit of fitting function. Moreover, taking advantage of the high resolution of these new simulation, we also explore the angular dependence of neutrino overdensity distributions around haloes. Neutrino density profiles have been obtained always assuming isotropy, but, in the surroundings of super-clusters, some preferred direction of motion could exist depending on the dark matter particles distribution. If neutrinos follow a particular direction, some asymmetry could be created. We investigate and quantify this effect for different kinds of clusters through the neutrino density profiles. In addition, we also compute neutrino density maps in the centre of haloes to confirm the neutrino wakes signatures proposed in~\cite{Zhu,LoVerde}.

This paper is organised as follows. In \S~\ref{sec:Methodology} we present the new HR-DEMNUni simulations that have been used to carry out all analysis presented in this work as well as the method used to compute the density profiles. \S~\ref{sec:profiles} gathers our main results, several density profiles for cold dark matter and neutrino particles in haloes with different masses. Some of these profiles take into account the angular distribution of the neutrinos around dark matter haloes. Finally, the discussion and conclusions are addressed in \S~\ref{sec:discussion}.

\section{Methodology}
\label{sec:Methodology}
In this section we describe the simulations employed and the method applied to compute density profiles, starting with the details of the high-resolution simulations in \S~\ref{sec:simulations} and the expressions for the profiles in \S~\ref{sec:kernel}.

\subsection{Numerical Simulations}
\label{sec:simulations}
In this analysis, we have used the ``Dark Energy and Massive Neutrino Universe'' (DEMNUni) suite of large N-body simulations~\cite{carbone_2016}. The DEMNUni simulations have been produced with the aim of investigating the large-scale structure of the Universe in the presence of massive neutrinos and dynamical DE, and they were conceived for the nonlinear analysis and modelling of different probes, including dark matter, halo, and galaxy clustering~\cite{castorina_2015,moresco_2016,zennaro_2018,ruggeri_2018,bel_2019,parimbelli_2021,parimbelli_2022,Guidi_2022, Baratta_2022, Gouyou_Beauchamps_2023, SHAM-Carella_in_prep}, weak lensing, CMB lensing, Sunyaev-Zel'dovich and integrated Sachs-Wolfe effects~\cite{carbone_2016,roncarelli_2015,fabbian_2018, Beatriz_2024}, cosmic void statistics~\cite{kreisch_2019,schuster_2019,verza_2019,verza_2022a, verza_2022b, Verza_etal_2024}, as well as cross-correlations among these probes~\cite{Vielzeuf_2022,Cuozzo2022}.

In particular, we have used the new high-resolution (HR) simulations with 64 times better mass resolution than previous standard runs:  the HR-DEMNUni simulations are characterized by a comoving volume of $(500 \: h^{-1}\mathrm{Gpc})^3$ filled with $2048^3$ dark matter particles and, when present, $2048^3$ neutrino particles. The simulations are initialized at $z_{\rm in}=99$ with Zel'dovich initial conditions. The initial power spectrum is rescaled to the initial redshift via the rescaling method developed in~\cite{zennaro_2017}. Initial conditions are then generated with a modified version of the \texttt{N-GenIC} software, assuming Rayleigh random amplitudes and uniform random phases. The HR-DEMNUni set consists of two simulations with total neutrino masses of $\sum m_\nu = 0,\, 0.16\, {\rm eV}$ considered in the degenerate mass scenario with three active neutrinos. The other cosmological parameters of the simulations are based on a Planck 2013~\cite{planck2013} LCDM reference cosmology (with massless neutrinos), in particular: $n_{\rm s}=0.96$, $A_{\rm s}=2.1265 \times 10^{-9}$, $h=H_0/[100\, 
{\rm km} \, s^{-1}{\rm Mpc}^{-1}]=0.67$, $\Omega_{\rm b}=0.05$, and $\Omega_{\rm m}=\Omega_{\rm CDM} + \Omega_{\rm b} + \Omega_\nu =0.32$; $H_0$ is the Hubble constant at the present time, $n_{\rm s}$ is the spectral index of the initial scalar perturbations, $A_{\rm s}$ is the scalar amplitude, $\Omega_{\rm b}$ the baryon density parameter, $\Omega_{\rm m}$ is the total matter density parameter, $\Omega_{\rm CDM}$ the cold dark matter density parameter, and $\Omega_\nu$ the neutrino density parameter. In the presence of massive neutrinos, $\Omega_{\rm b}$ and $\Omega_{\rm m}$ are kept fixed to the above values, while $\Omega_{\rm CDM}$ is changed accordingly. Tab.~\ref{tab:neutrino_params} summarizes the masses of the CDM and neutrino particles together with the neutrino fraction $f_\nu \equiv \Omega_\nu / \Omega_{\rm m}$. 
\begin{table}[t]
\centering
\vspace{2ex}
\setlength{\tabcolsep}{0.7em}
\begin{tabular}{cccc}
\toprule
\multicolumn{4}{c}{HR-DEMNUni} \\
\midrule
$\sum m_\nu$  [eV] & $f_\nu$ & $m_{\rm p}^{\rm CDM}$  [\hmsun] & $m_{\rm p}^\nu$  [\hmsun] \\
\hline
0  & 0 & $1.2921\times 10^{9}$ & $0$ \\
\hline
0.16  & 0.012 & $1.2767\times 10^{9}$ & $1.5441\times 10^7$ \\
\bottomrule
\end{tabular}
\caption{
Summary of particle masses and neutrino fractions implemented in the HR-DEMNUni simulations. The first column shows the total neutrino mass, the second the fraction of neutrinos and matter density parameters, and the last two columns show the corresponding mass of CDM and neutrino particles implemented in the simulations. 
}
\label{tab:neutrino_params}
\end{table}
Dark matter haloes are identified using a friends-of-friends (FoF) algorithm~\cite{davis_1985_fof}, with a linking length of 0.2 times the mean particle separation, applied to both CDM and neutrino particles in the simulation, with a minimum number of particles per species fixed to 32, corresponding to a mass of $\sim 4 \times 10^{10} h^{-1}M_{\odot}$. FoF haloes are further processed with the {\sc subfind} algorithm~\cite{springel_2001_gadeget,dolang_2009_gadget} to produce subhalo catalogues. With this procedure, some of the initial FoF parent haloes are split into multiple substructures. In particular, in this work we use the spherical overdensity halo catalogues, so-called $M_{\rm 200b}$, identified by the {\sc subfind} algorithm, and in the following we will refer to them with the term ``halo''.

\subsection{Computing Density Profiles}
\label{sec:kernel}
In order to compute smoothed density profiles we use the kernel-based algorithm proposed in~\cite{kernel}. This density estimator is based on a variable window width. 
\begin{equation}
    \hat{\nu}(r)=\sum_{i=1}^{N}\frac{1}{h_i^3}\tilde{K}(r,r_i,h_i),
\end{equation}
where $h_i$ is the window width associated with the $i$th particle and $\tilde{K}$, a Gaussian kernel. 
\begin{equation}
    \tilde{K}(r,r_i,h_i) = \frac{1}{2(2\pi)^{3/2}}\left(\frac{rr_i}{h_i^2}\right)^{-1}\left[e^{-(r_i-r)^2/2h_i^2}-e^{-(r_i+r)^2/2h_i^2}\right]
\end{equation}
The window width varies with position in such a way that the bias-to-variance ratio of the estimate is relatively constant. We chose the window width in the same way as~\cite{kernel}, $h_i\propto r^{1/2}$ is set at $0.1r_{vir}$ as $h_{0.1}=0.05r_{vir}$. We use this kernel to compute all density profiles in this paper.

\section{Density profiles}
\label{sec:profiles}
In this Section we present our main results. Firstly, we will focus on density profiles integrated in angles both for cold dark matter and neutrino distributions within haloes of different masses, studying if new density profiles measured from the HR-DEMNUni simulations fit the formulas proposed by~\cite{Villaescusa-Navarro_2013} with the aim to reproduce them. Secondly, we will consider how density profiles could change depending on the neutrino direction and compute them in solid angles ahead of and behind the halo centres along the direction of motion of cold dark matter particles. Lastly, we will look at the centre of the selected haloes and look for backward neutrino wakes as those proposed in~\cite{Zhu,LoVerde}. All analyses have been carried out over a redshift zero realization of the HR-DEMNUni simulations.

\subsection{Profiles as a function of radius}
\label{sec:profiles_radius}
\begin{figure}[h!]
    \centering
    \includegraphics{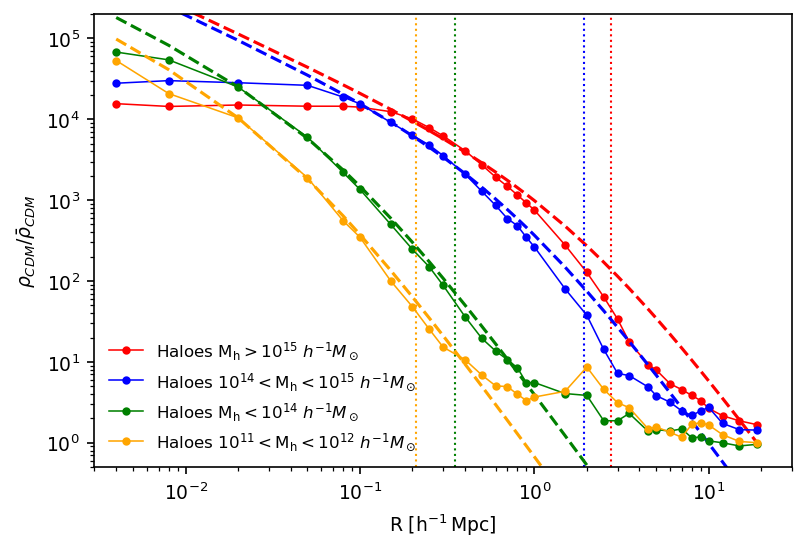}
    \caption{CDM overdensity profiles in haloes with masses $M_{\rm h}$ in different mass bins. Dashed lines correspond to a NFW density profile fit. Vertical dotted lines set the halo radii $\rm{R}_{200}$ defined by an overdensity threshold $\Delta=200$ with respect to the mean background total matter density $\rho_b$.}
    \label{fig:CDM_den_prof}
\end{figure}

As a first check, we have calculated the CDM density profiles for four different halo masses at redshift zero (Figure~\ref{fig:CDM_den_prof}) and fit them to the standard NFW profile (dashed lines in Figure~\ref{fig:CDM_den_prof}). For these and the following fits, the \texttt{python} package \texttt{lmfit}\footnote{\url{https://lmfit.github.io/lmfit-py/}} based on non-linear least-squares minimisation has been used.

The outcome is as expected, we see the plateau in the innermost regions, the higher the halo mass is the larger is this plateau, then the profiles drop quickly to the mean density around the virial radius and inside and outside over-densities differ by several orders of magnitude. In addition, and as it is already known, NFW profiles just return a good fit around the virial radius. These profiles are well known and have been calculated several times through literature, we use them here just to show that the algorithm used to compute the density profiles works correctly, so it can be applied in the calculation of neutrino profiles. 

\begin{figure}[b]
    \centering
    \begin{tabular}{cc}
    \includegraphics[width=0.48\textwidth]{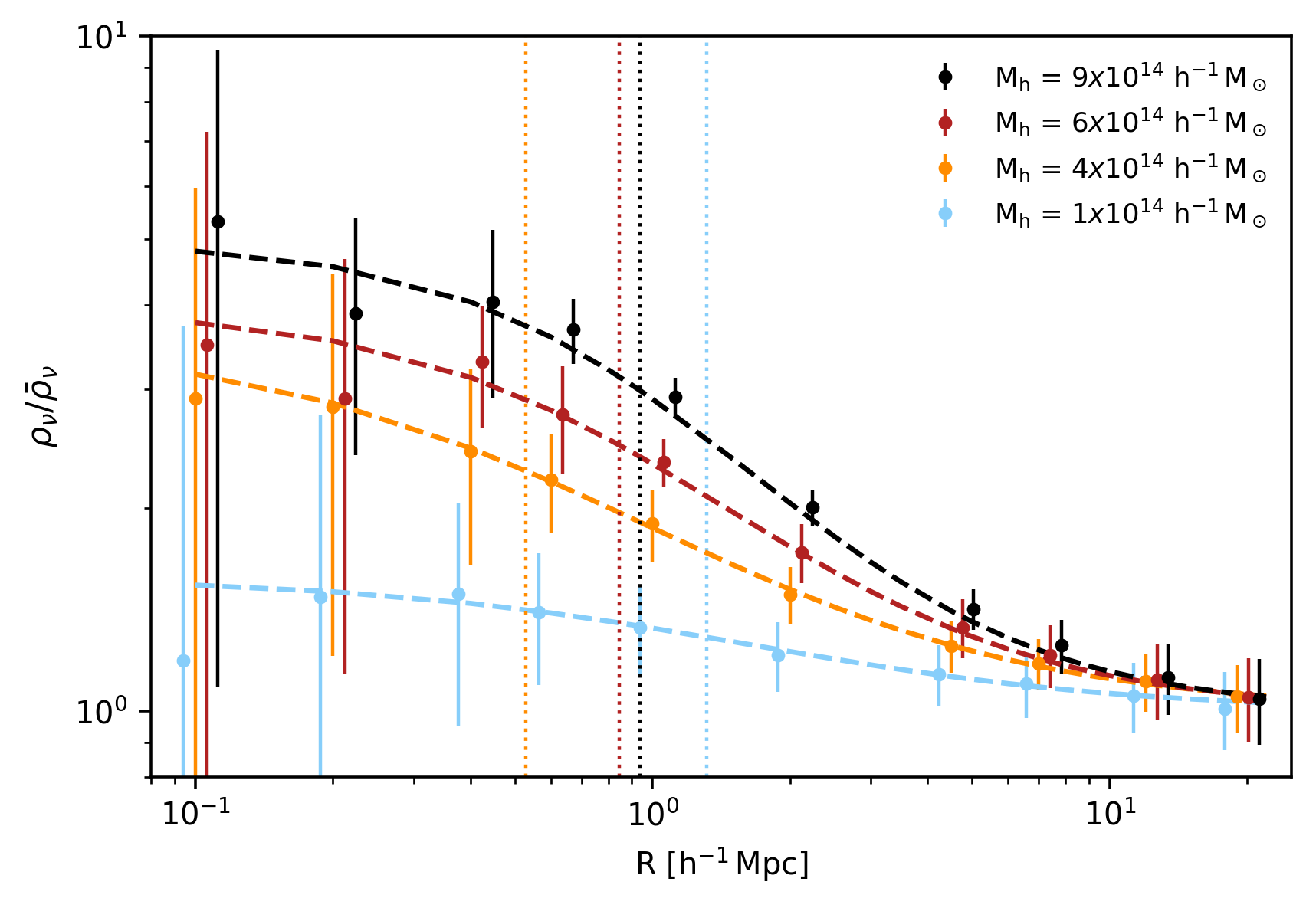} & \includegraphics[width=0.48\textwidth]{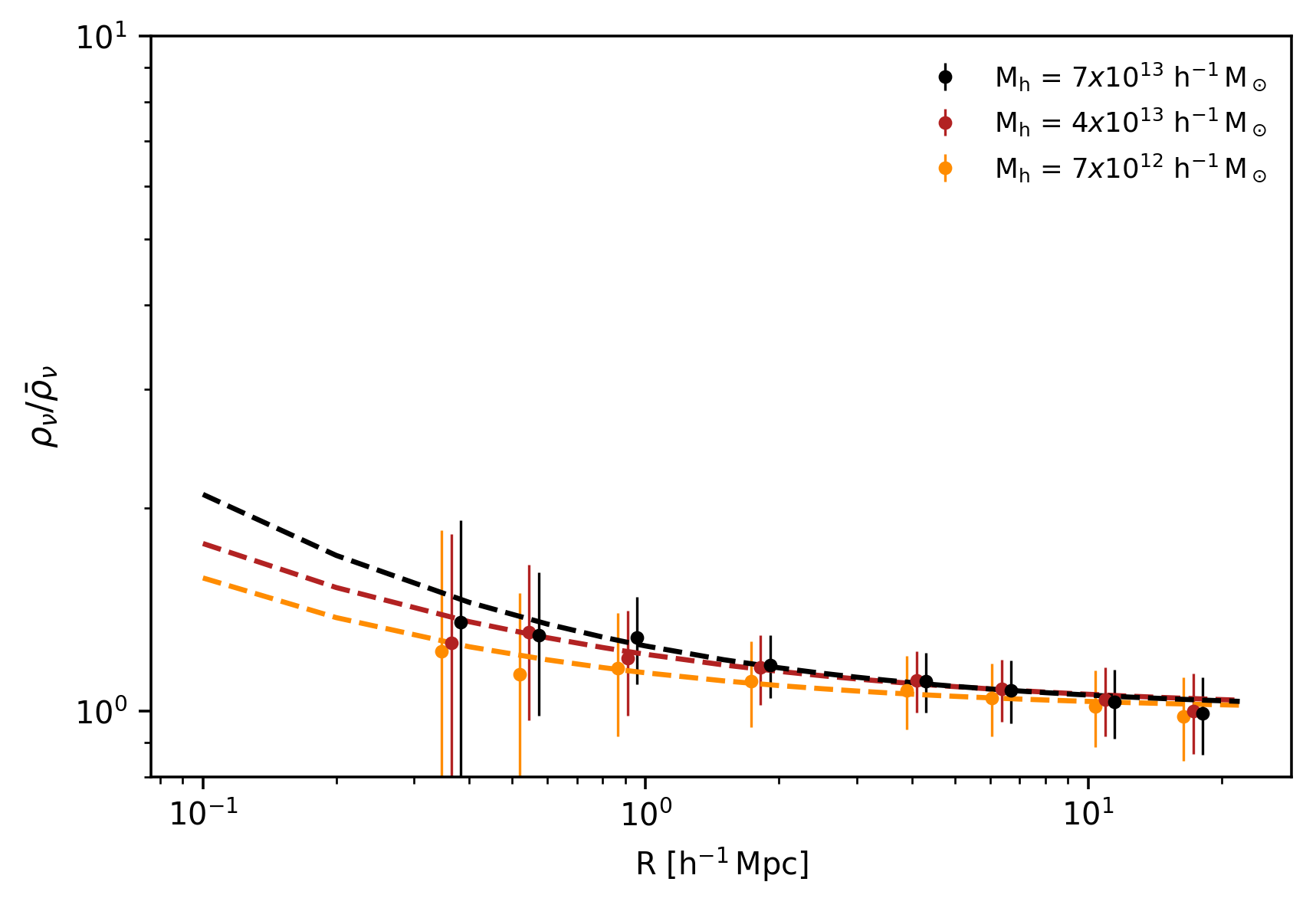}
    \end{tabular}
    \caption{Mean neutrino overdensity profiles for different halo masses. For each mass we select a maximum of $200$ haloes randomly selected within a mass bin of width $\pm5\%$ of the central value displayed in the legend. The error bars correspond to $1\sigma$ dispersion around the calculated mean value of the overdensity at each radius. LEFT: Neutrino overdensity profiles for $M_{\rm h} > 10^{14}\,h^{-1}M_\odot$.  In dashed lines the fits to formula~\eqref{eq:complex_fit}. Vertical dotted lines set the core radius ($r_c$ parameter) obtained from the fit. RIGHT: Mean neutrino overdensity profiles for $M_{\rm h} < 10^{14}\,h^{-1}M_\odot$. In dashed lines the fit to formula~\eqref{eq:simple_fit}.}
    \label{fig:nu_profile_masses}
\end{figure}

Now, we repeat the same procedure for the neutrino component, also at redshift zero, but in smaller halo mass bins. We plot the overdensity for neutrinos and use the formulas~\eqref{eq:nu_fitting_formulas} below to fit the profiles. We present the results for haloes with masses greater than $10^{14}\,h^{-1}M_\odot$ in the left panel of Figure~\ref{fig:nu_profile_masses}, along with the results of the fit to Equation~\eqref{eq:complex_fit} which is plotted as dashed lines. Only some of them are displayed in the left panel of Figure~\ref{fig:nu_profile_masses}  for ease of visualisation, but all fit results are included in Table~\ref{tab:complex_fit}. Likewise, in the right panel of Figure~\ref{fig:nu_profile_masses}, the density profiles obtained for haloes with masses between $10^{11}\, h^{-1}M_\odot \le M_{\rm h} \leq 10^{14} h^{-1}M_\odot$ are shown combined with the fitting result to Equation~\eqref{eq:simple_fit}, plotted as dashed lines too. As well, only a few are displayed in the right panel of Figure~\ref{fig:nu_profile_masses} but all results are included in Table~\ref{tab:simple_fit}. Due to computational limitations, the angle-averaged profiles, corresponding to each of the halo masses listed in Tables~\ref{tab:complex_fit} and \ref{tab:simple_fit}, has been calculated by averaging over neutrino profiles measured from 200 haloes. All available haloes has been used for masses close to $10^{15}$\hmsun. In Figure~\ref{fig:nu_profile_masses} the error bars represent the dispersion \textit{around} the mean density profile, rather than the dispersion \textit{in} the mean density profile as instead assumed in~\cite{Villaescusa-Navarro_2013}. So our errors are more conservative and mildly depend on the number of analysed haloes. We performed our fits using the equations below:
\begin{subequations}
\begin{alignat}{2}
\delta_\nu(r) =&\frac{\rho_\nu(r)-\bar{\rho}_\nu}{\bar{\rho}_\nu} = \frac{\rho_c}{1+(r/r_c)^\alpha} \hspace{1cm} & \text{for halo masses M}>10^{14}\,h^{-1}M_\odot \label{eq:complex_fit}\\
\delta_\nu(r) =& \kappa/r^\alpha & \text{for halo masses M}<10^{14}\,h^{-1}M_\odot \label{eq:simple_fit}
\end{alignat}
\label{eq:nu_fitting_formulas}
\end{subequations}

\begin{table}[t!]
    \centering
    \begin{tabular}{c|c c c}
         $M_{\rm h}\ [h^{-1}M_\odot]$ & $\rho_c $ & $r_c \ [h^{-1}\rm{Mpc}]$ & $\alpha$\\
         \hline
         $1\times10^{14}$ & $0.58 \pm 0.12$ & $1.31 \pm 0.57$ & $1.04 \pm  0.16$ \\
$2\times10^{14}$ & $1.23 \pm 0.21$ & $0.68 \pm 0.22$ & $1.05 \pm  0.09$ \\
$4\times10^{14}$ & $2.55 \pm 0.30$ & $0.53 \pm 0.11$ & $1.03 \pm  0.05$ \\
$6\times10^{14}$ & $2.96 \pm 0.33$ & $0.85 \pm 0.16$ & $1.25 \pm  0.08$ \\
$8\times10^{14}$ & $3.58 \pm 0.44$ & $0.92 \pm 0.18$ & $1.30 \pm  0.10$ \\
$9\times10^{14}$ & $3.98 \pm 0.33$ & $0.94 \pm 0.12$ & $1.38 \pm  0.07$ \\
    \end{tabular}
    \caption{Coefficients of Equation~\eqref{eq:complex_fit} fitted to the mean neutrino overdensity profiles. Results for neutrinos within haloes with $M_{\rm h} > 10^{14}$\hmsun in HR-DEMNUni simulations. For each mass we analyze a maximum of $200$ haloes randomly selected within a mass bin of width $\pm5\%$ of the central value displayed in the first column.}
    \label{tab:complex_fit}
\end{table}
\begin{table}[t!]
    \centering
    \begin{tabular}{c|c c}
         $M_{\rm h}\ [h^{-1}M_\odot]$ & $\kappa$ & $\alpha$\\
         \hline
$4\times10^{11}$ & $0.140\pm0.020$ & $0.62\pm0.12$ \\
$7\times10^{12}$ & $0.140\pm0.020$ & $0.61\pm0.13$ \\
$1\times10^{13}$ & $0.136\pm0.018$ & $0.53\pm0.11$ \\
$4\times10^{13}$ & $0.215\pm0.020$ & $0.56\pm0.08$ \\
$7\times10^{13}$ & $0.250\pm0.022$ & $0.64\pm0.09$ \\
    \end{tabular}
    \caption{Coefficients of Equation~\eqref{eq:simple_fit} fitted to the mean neutrinos overdensity profiles. Results for neutrinos within haloes with $M_{\rm h} < 10^{14}$\hmsun in the HR-DEMNUni simulations. For each mass we analyze a maximum of $200$ haloes randomly selected within a mass bin of width $\pm5\%$ of the central value displayed in the first column.}
    \label{tab:simple_fit}
\end{table}

In~\cite{Villaescusa-Navarro_2013}, the authors found that Equations~\eqref{eq:nu_fitting_formulas} described very well the average neutrino overdensity profiles over a wide range of radii. The physical meaning of the parameters in the profile~\eqref{eq:complex_fit} is very simple: $r_c$ and $\rho_c$ represent the length and the overdensity of the core in the overdensity profile of the neutrino halo, while $\alpha$ is a parameter that controls how fast the overdensity profile falls on large radii. In~\cite{Villaescusa-Navarro_2013} it was also claimed that for haloes with mass below $\sim10^{13.5}\,h^{-1}M_\odot$ the resolution in their N-body simulations was not large enough to properly resolve the core in the neutrino density profiles, so Equation~\eqref{eq:simple_fit} was proposed to reproduce the outskirts of the neutrino density profiles. In our case, even though our simulations have much larger mass resolution, the neutrino mass is smaller than in~\cite{Villaescusa-Navarro_2013}  and we can not resolve the core either, if any core exists for such small neutrino masses. So we use the same expression~\eqref{eq:simple_fit} to fit the neutrino profiles in haloes with masses below $10^{14}\,h^{-1}M_\odot$. It is important to emphasize that in our case the small neutrino mass, therefore the high thermal neutrino velocity, is preventing the neutrino clustering in haloes with masses less than $10^{14}\,h^{-1}M_\odot$, which results in no neutrino core formation.

As the halo mass becomes smaller, the neutrino overdensity profile becomes flatter, meaning a larger neutrino clustering in more massive haloes. Comparing the neutrino profiles in the left and right panels of Figure~\ref{fig:nu_profile_masses} with the CDM overdensity profiles in Figure~\ref{fig:CDM_den_prof}, it is possible to observe that neutrino profiles are more extended than CDM ones, due to the larger CDM particles clustering as compared to hot neutrinos. If we compare neutrino profiles for large and low halo masses, we see a clear overdensity decrease of around a factor of two in the latter case, which is a consequence of the decrease of the halo mass and the associated gravitational potential well. Paying attention to the tails of the neutrino profiles, we can verify that they are well reproduced by the proposed fitting functions. At this point, it is worth noting that formulas (\ref{eq:nu_fitting_formulas}), which were proposed studying simulations with larger neutrino masses, work reasonably well also for these new simulations where the total neutrino mass is twice smaller.

\begin{figure}[t!]
    \centering
    \includegraphics[width=0.95\textwidth]{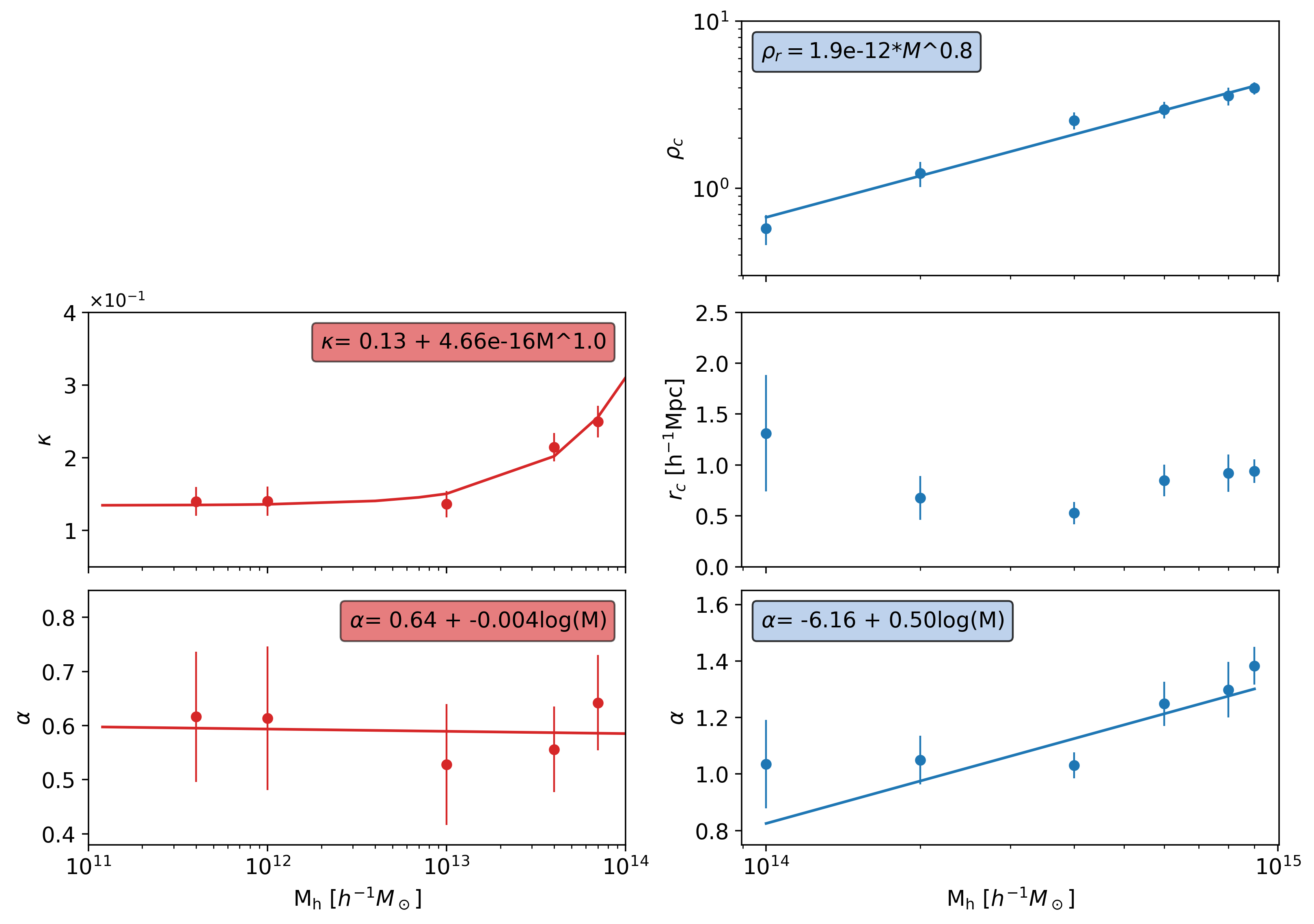}
    \caption{Derived parameters of the fitting functions~\ref{eq:complex_fit} (right) and~\ref{eq:simple_fit} (left) used to characterize the neutrino overdensity profiles. In both columns we show the dependence of the respective parameters as a function of the masses of their host halo.}
    \label{fig:parameters_fit}
\end{figure}

In Figure~\ref{fig:parameters_fit}, we show the parameters derived from the fitting functions against the mass of the halo hosting the neutrino halo for the two cases of neutrino profiles~\eqref{eq:complex_fit}-\eqref{eq:simple_fit}, with the aim to check if the trend is the same as in~\cite{Villaescusa-Navarro_2013}. We find that our results for the $\alpha$ parameters are quite comparable and, in both cases, the tendency is alike to that obtained in~\cite{Villaescusa-Navarro_2013}. The slopes of the fitting functions are close to zero, suggesting that there is not much mass dependence for the two $\alpha$ parameters. Regarding the core density parameter, $\rho_c$, the trend we find is comparable as well: the density of the neutrino core increases with the mass of the host halo, nevertheless the results of the fit are slightly different than in~\cite{Villaescusa-Navarro_2013}. We get a fit for the $\kappa$ parameter: its trend is a some constant regime from $10^{11}$ up to $1\times10^{13}\,h^{-1}M_\odot$ and then it starts to grow. On distances much larger than the core radius, $r\gg r_c$, the profile \eqref{eq:complex_fit} reduces to \eqref{eq:simple_fit}, with $\kappa = \rho_cr_c^\alpha$, so we expected a decreasing of $\kappa$ as the halo mass decrease and so it is. Particularly relevant is the trend observed for $r_c$ for which we do not get a fit, as shown in the $r_c$ panel of Figure~\ref{fig:parameters_fit}. In~\cite{Villaescusa-Navarro_2013} $r_c$ grows with the halo mass, but what we have found from our analysis is that it decreases with the halo mass for $M_{\rm h} \lesssim 4 \times 10^{14}$\hmsun, while it starts to grow for $M_{\rm h} > 4 \times 10^{14}$\hmsun, recovering a similar trend as in~\cite{Villaescusa-Navarro_2013} (see Table \ref{tab:complex_fit}). As the expected tendency, also considering what happens for the CDM core, is that the radius of the neutrino core grows with the mass of the host halo, this result suggest that Equation~\eqref{eq:complex_fit} may not be applicable for haloes with masses lower than $4 \times10^{14}$\hmsun. A higher resolution simulation and, especially, a lower total neutrino mass seem to have pushed the limit: for haloes with $M_{\rm h} < 4 \times10^{14}$\hmsun the obtained profiles are too flat for formula~\eqref{eq:complex_fit} to resolve the core, so the core information is washed out. 

This could mean that, for neutrino total masses less than $\sim 0.16$ eV and CDM halo masses less than a few $10^{14}$\hmpc, neutrinos behave more as light-like particles and this prevents neutrino core formation. In~\cite{Villaescusa-Navarro_2013}, their results suggest that the fitting formula does not work for haloes with masses less than $10^{13.5}$\hmsun, now, for less massive neutrinos we put the limit higher suggesting $4 \times10^{14}$\hmsun. It is possible that the limit can be push higher, maybe around $10^{15}$\hmsun with N-body simulations implementing total neutrino masses lower than $0.16$ eV. 

Following the previous argument, such a low neutrino mass produces a very flat profiles with very large error bars in the innermost halo regions, so a limit, where we are dominated by the dispersion of the profiles, is reached meaning that no reliable information can be obtained in these regimes.

Our simulations thus predict what averaged neutrino profiles should look like in the presence of CDM, although some new limits have been found. In order to obtain more conclusive results, more high-resolution simulations like the one used in this work with different and smaller neutrino masses are needed. Nevertheless, we can conclude that, if the signal is measured by future weak lensing surveys, this should provide a clear signature that massive neutrinos do exist.

\subsection{Profiles as a function of angle}
\label{sec:profiles:angle}
It has always been assumed that neutrino haloes, produced in areas with a high dark matter density, are spherical; but if neutrinos fall in those areas because of dark matter, the CDM distribution could produce a preferential direction of neutrino in-falling which would break the isotropy, resulting in some kind of ovoid neutrino haloes. For some dark matter haloes, we expect a deformed associated neutrino halo that would present a higher neutrino density in the front part in the direction of motion of the CDM and a lower neutrino density behind the halo centre along the direction of motion as well. 

In order to search for this effect, and as we want to see an effect on the neutrino density profiles as a function of the angle, a reference direction, with respect to which the angles are measured, is needed. We set the reference direction for each analysed halo as the direction resulting from averaging the velocity vectors of all CDM particles within the halo, assuming that the haloes extend up to 20 $h^{-1}$Mpc, since neutrino haloes have much larger radii than the corresponding CDM haloes. Then, we compute the neutrino overdensity profiles ahead of and behind the halo centre along this reference direction. 

In practice, the procedure followed is: first, find the average CDM particle velocity in the halo, then rotate all the system, CDM and neutrino particles (up to a distance of 20 $h^{-1}$Mpc radius from the center of the halo) so that the average halo CDM velocity is aligned to the x-axis, i.e. $\theta=90^\circ$ and $\phi=0^\circ$ in spherical coordinates; then, the neutrino position distribution, after the rotation, is computed. 

Another anisotropy effect is the one proposed in~\cite{Zhu,LoVerde}. This produces neutrino wakes in the inner regions of the neutrino halo due to the peculiar motion of the halo itself. In order to investigate such an effect we look at the centre of the halo in a box of 6 $h^{-1}$Mpc side.

\subsubsection{Front loading neutrinos}
\label{sec:front loading}

\begin{figure}[b!]
    \centering
    \includegraphics[scale=0.9]{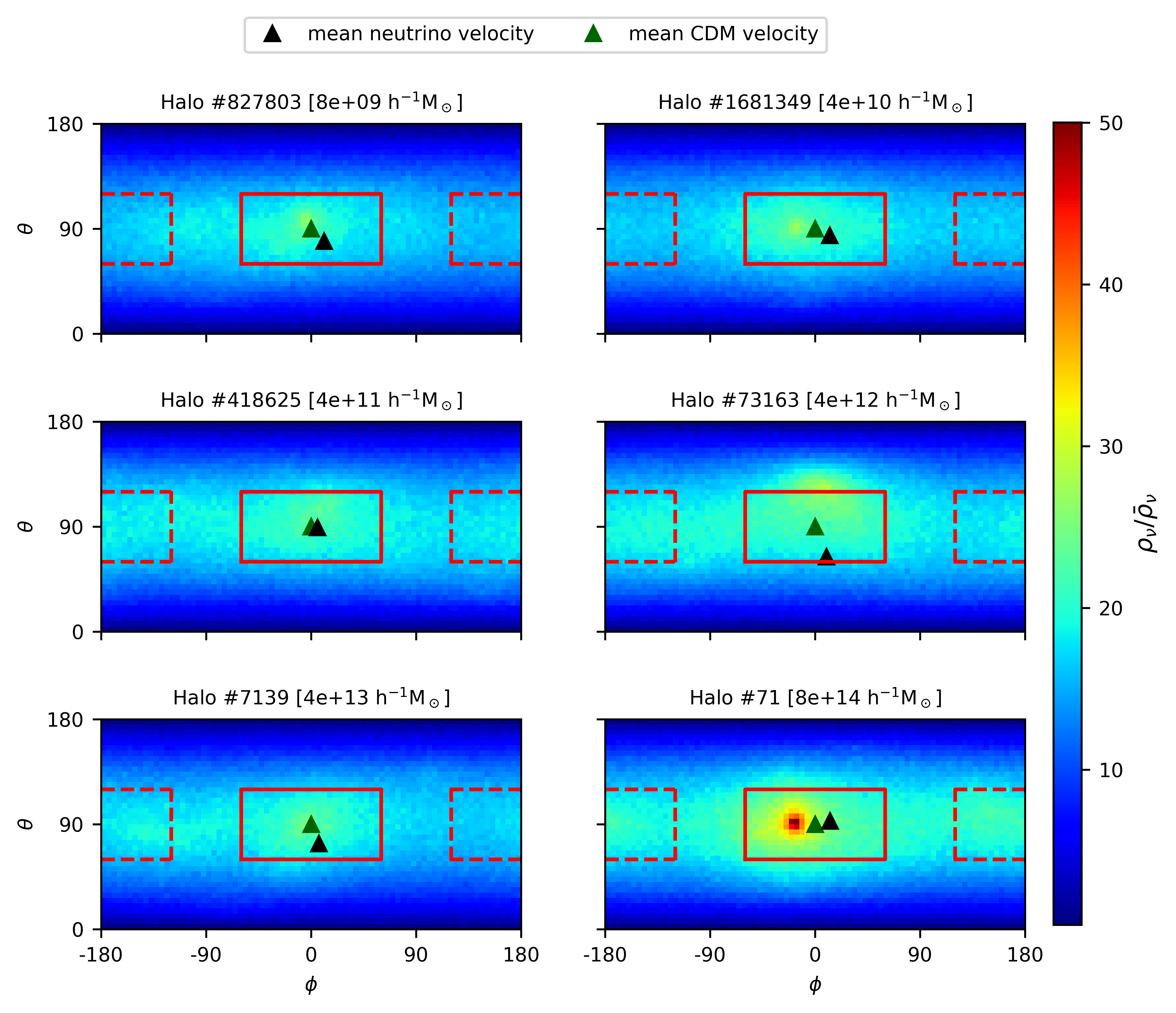}
    \caption{Neutrino overdensity in spherical coordinates for some selected haloes of different masses. Triangle markers represent mean CDM and neutrino velocity vector in spherical coordinates. Red solid line marks the neutrinos ahead of the halo centre in the direction of the CDM motion and red dashed line marks neutrinos behind.}
    \label{fig:pos_selection}
\end{figure}

As noted above, neutrinos falling in a particular halo will see a preferential direction, i.e. the average CDM direction of motion, and will fall following it. Following that direction, the closer the neutrinos travel to the halo centres, the more gravitational force they will suffer and the more they will deviate from their trajectory towards the halo central axis. As a result, a region of overdensity will form ahead of the halo centres in the direction of CDM motion.

In practice, if neutrinos are to cluster further ahead of the centre of the halo, the expected mean angle values are $\theta = 90^\circ$ and $\phi = 0^\circ$ for construction, because we have rotated each halo system such that the CDM mean velocity vector is aligned to the x-axis ($\theta = 90^\circ$ and $\phi = 0^\circ$, in spherical coordinates). To analyze this effect, we have randomly selected a total of 120 typical haloes of different masses. In all of these haloes we observe that the largest clustering occurs ahead of the halo centre in the direction of motion. Some examples are shown in Figure~\ref{fig:pos_selection}.

\begin{figure}[b!]
    \centering
    \includegraphics[width=0.98\textwidth]{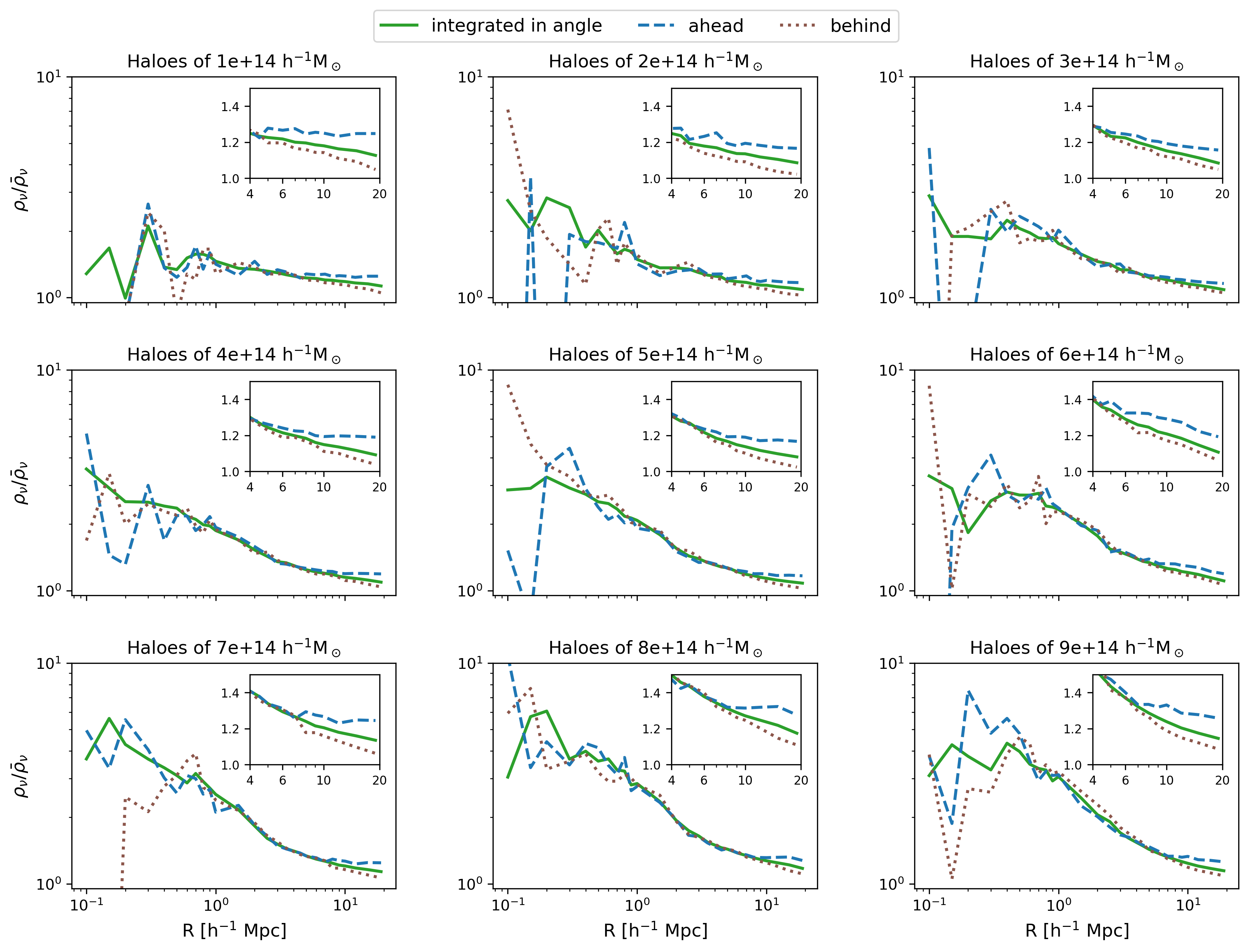}
    \caption{Neutrino profiles for haloes with masses between $1\times10^{14}$ and $9\times10^{14} \ h^{-1}M_\odot$. Solid line represent profiles integrated in angles. Dashed line show profiles obtained from neutrinos located ahead of the halo centre in the direction of the CDM motion. Dotted line show profiles obtained from neutrinos located behind the halo centre in the direction of CDM motion.}
    \label{fig:asym_profiles_halos}
\end{figure}

\begin{figure}[t!]
    \centering
    \includegraphics[width=0.98\textwidth]{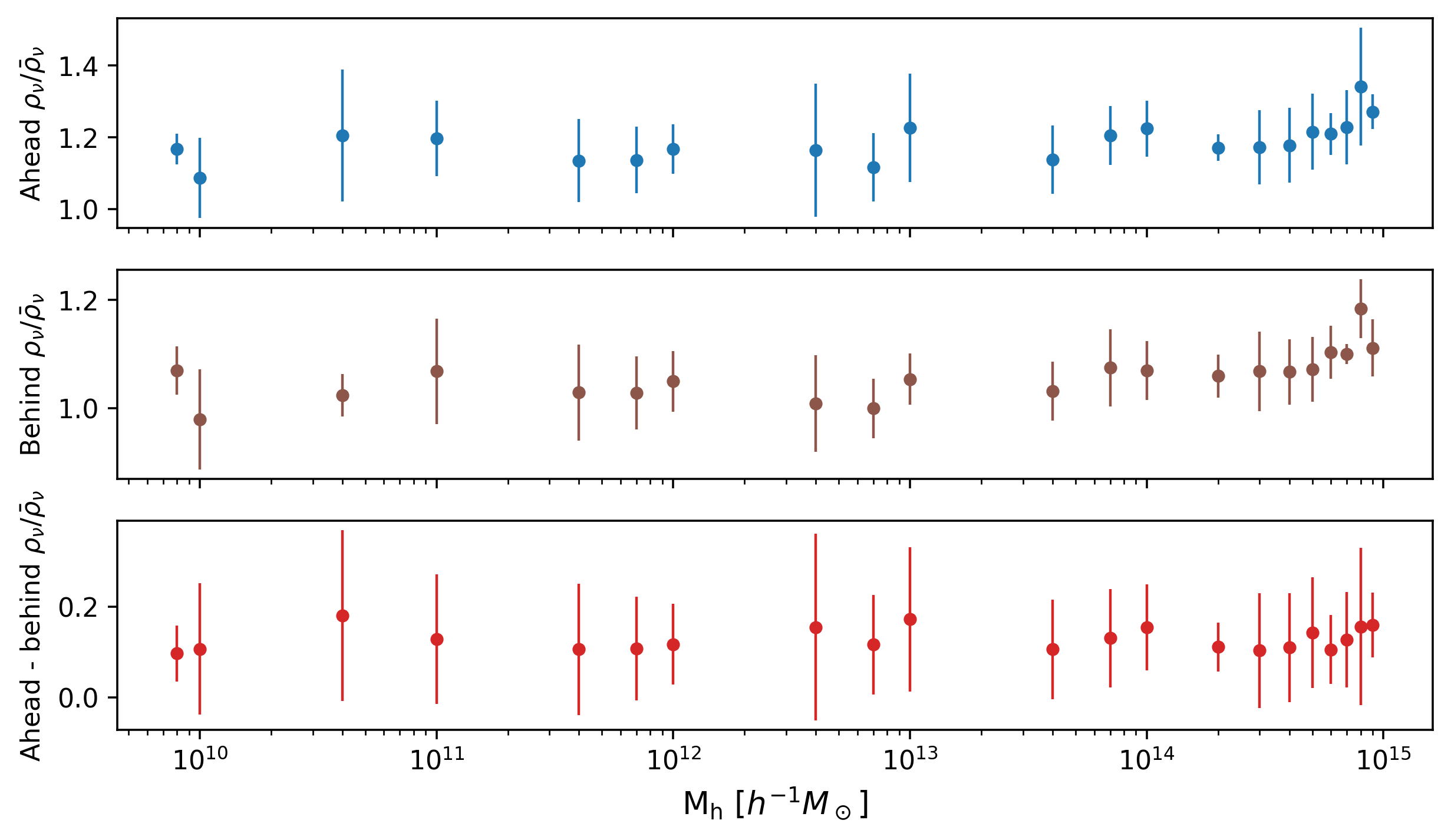}
    \caption{Neutrino overdensity for neutrinos located ahead of and behind the halo centre in the direction of the CDM motion as a function of the dark matter halo masses. The difference between ahead overdensity and behind overdensity means is plotted in the bottom subplot. The error bars represent $1\sigma$ error of the mean values.}
    \label{fig:asymmetry_mass}
\end{figure}

From the simulations we see that the neutrino mean velocity is more or less aligned to the CDM mean velocity for each analyzed halo, as shown in Figure~\ref{fig:pos_selection}, where black triangles represent the CDM mean velocity and green triangles the neutrino mean velocity. This was the first lineal effect studied in the cosmic neutrino background framework. It can also be observed how neutrinos cluster along the direction of CDM motion ($45^\circ<\theta<135^\circ$) and how they cluster more in the areas located ahead of the halo centres ($\theta=90^\circ$ and $\phi=0^\circ$) marked by a red solid line square, and less in the areas located behind the halo centres ($\theta=90^\circ$ and $\phi=180^\circ$) marked by a red dashed line square. The effect is present in all halo masses and its amount is not so different among the analysed haloes (see Figure~\ref{fig:asymmetry_mass}). This effect depends on the CDM distribution of the halo and on the haloes surrounding it; actually, we made an extended analysis for about hundreds of haloes and selected those where the effect is larger, as the ones shown in Figure~\ref{fig:pos_selection}. The results we present in this work correspond to an analysis of a total number of 120 haloes of different masses.

Using the selected haloes, we compute the neutrino overdensity profiles, similarly to the previous ones in the left and right panels of Figure~\ref{fig:nu_profile_masses}, but now without averaging on the $4\pi$ solid angle. We select a region defined by $\Delta\theta=60^\circ$ and $\Delta\phi=120^\circ$ with respect to the centre placed at $\theta=90^\circ$ and $\phi=0^\circ$ in order to catch neutrinos ahead the halo centre, and with centre placed at $\theta=90^\circ$ and $\phi=180^\circ$ in order to catch neutrinos behind the halo centre in the direction of the CDM flow (see the squares drawn in Figure~\ref{fig:pos_selection}). Then, we calculate the profiles of these neutrino subsets. The results for haloes with masses between $1\times10^{14}$ and $9\times10^{14}\,h^{-1}M_\odot$, together with the profile integrated in angles, are shown in Figure~\ref{fig:asym_profiles_halos}. The effect becomes noticeable beyond a radius of  5 $h^{-1}$Mpc  where the three lines diverge: the dashed one, which corresponds to neutrinos ahead, increases and the dotted one, which corresponds to neutrinos behind, decreases, while the solid one, which corresponds to all neutrinos within the halo, stays in between.

To quantify the asymmetry effect we have integrated the profiles from radius equal to 6 up to 20 $h^{-1}$Mpc, for both ahead and behind profiles. The result is an overdensity asymmetry above $0.1$ (Figure~\ref{fig:asymmetry_mass}) which remains constant over the dark matter halo mass and slightly increases for masses close to $10^{15}\,h^{-1}M_\odot$.

\subsubsection{Wakes in neutrino haloes}
\label{sec:nu wakes}

\begin{figure}[t!]
    \centering
    \includegraphics[scale=0.8]{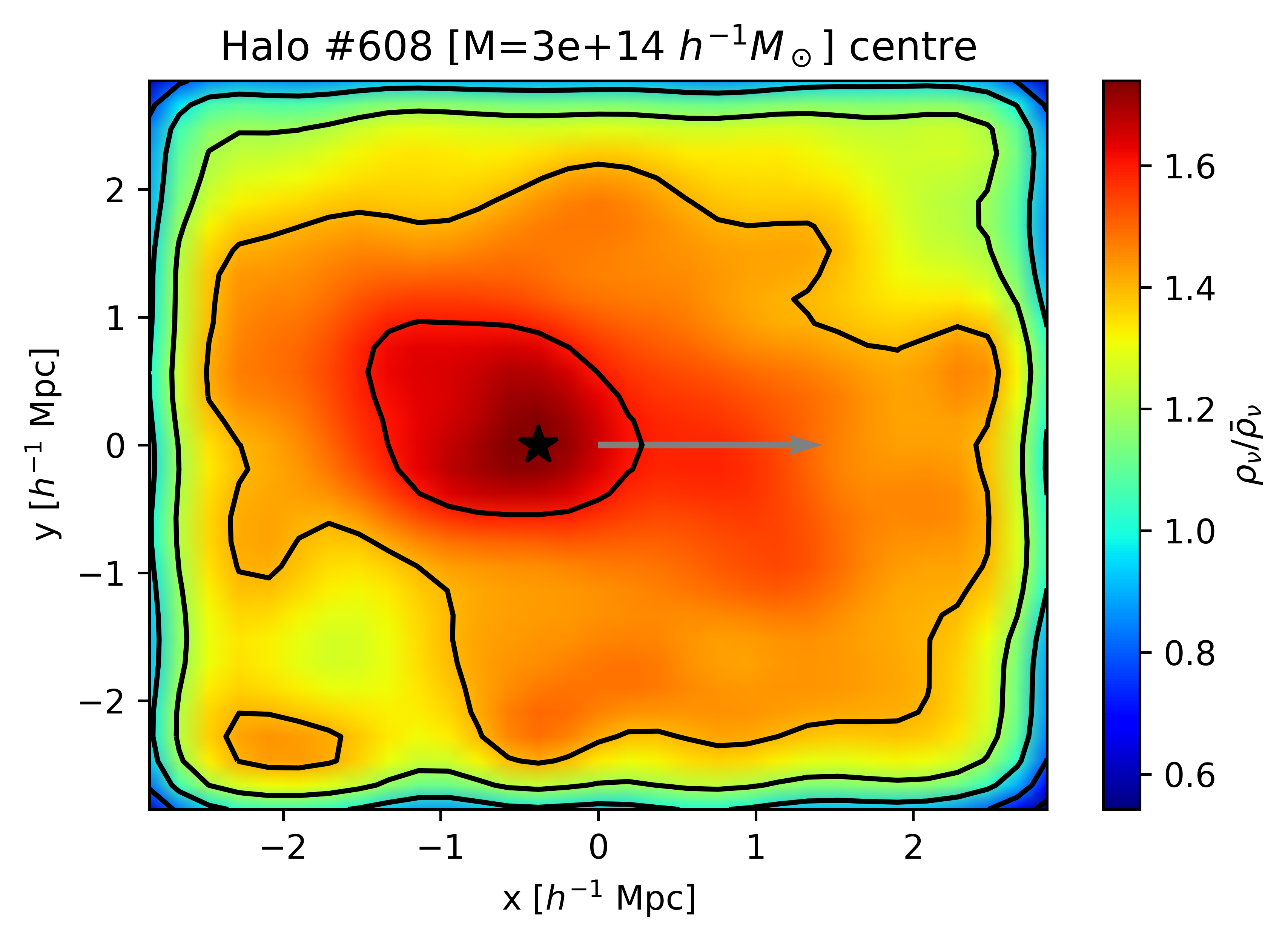}
    \caption{Neutrino overdensity distribution in the centre of one halo of mass $3\times10^{14}\,h^{-1}M_\odot$. The black star represents the position where the maximum value of the overdensity distribution is reached. The grey arrow shows the average direction of CDM motion.}
    \label{fig:pos_centre}
\end{figure}
\begin{figure}[t!]
    \centering
    \includegraphics[scale=0.6]{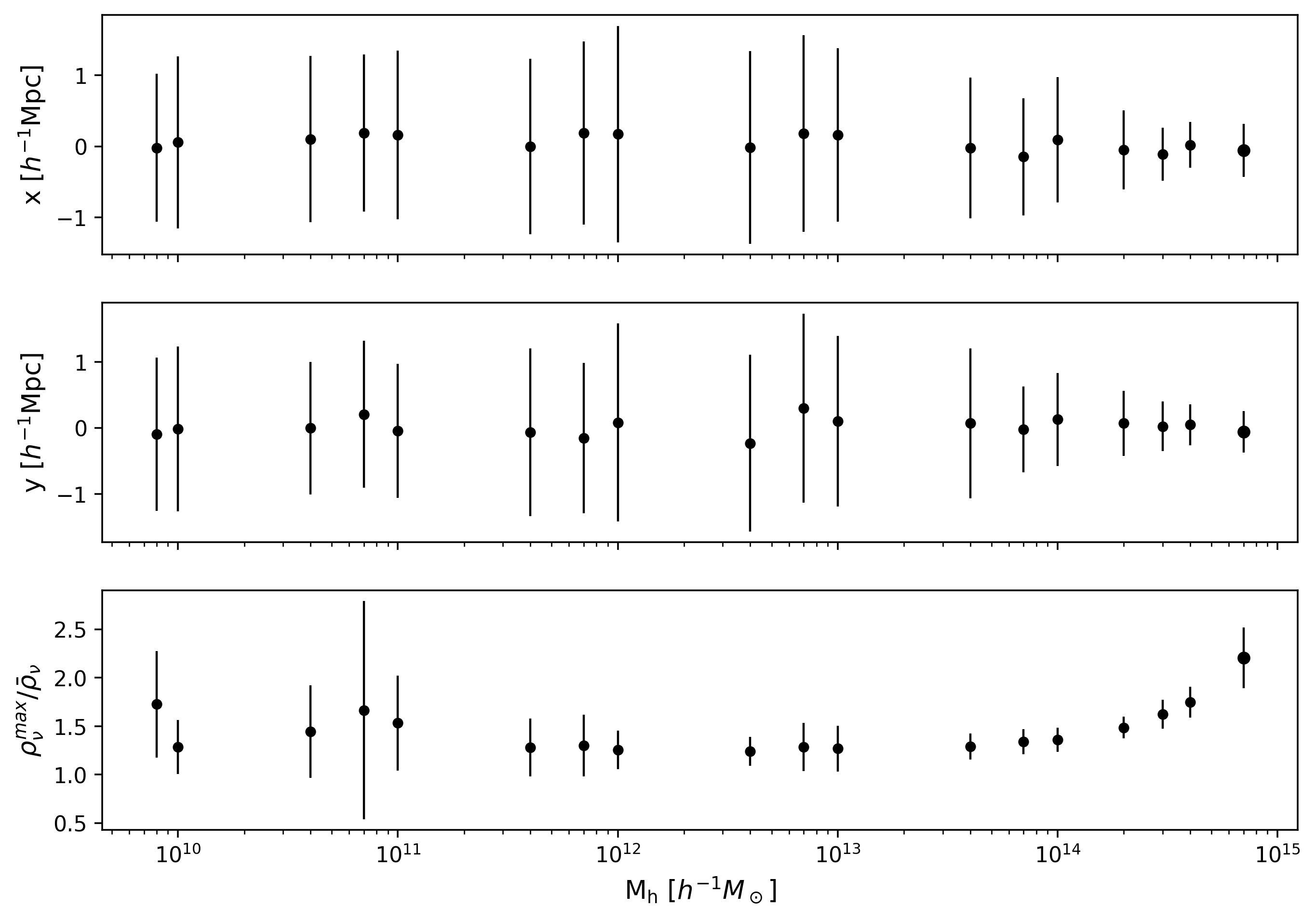}
    \caption{Mean x and y position of the maximum value of the 2D neutrino overdensity distribution in a box of 6 $h^{-1}$Mpc side centred on the centre of the halo as a function of the mass of the host halo. In the third subplot, the maximum value of the neutrino overdensity is plotted. All error bars correspond to $1\sigma$ error of the respective value.}
    \label{fig:pos_centre_array}
\end{figure}

In~\cite{Zhu,LoVerde}, it was shown that the peculiar motion of haloes causes neutrino particles to accumulate behind the moving halo, generating wakes that slow the halo down due to dynamical friction, and reducing the cross-correlation between neutrinos and CDM. Here, using the HR-DEMNUni simulations we look for these wakes at the centre of the haloes. To this aim, we use the same procedure as before: find the halo, determine the CDM mean velocity vector and rotate all the system such that the CDM mean velocity vector matches the x-axis at the centre of the halo. For each rotated halo, we select a box of 6 $h^{-1}$Mpc side centered at the halo centre and calculate the neutrino density distribution over xy-axis. This time, neutrino overdensity distributions are computed using a standard Gaussian kernel. One example of the final result is shown in Figure~\ref{fig:pos_centre}, where the black star marks the position of the maximum in the overdensity distribution. This plot is a clear example of the effect, some wakes can be observed and the maximum of the overdensity is displaced behind the centre along the CDM velocity direction which is illustrated by the grey arrow. 

We repeated this analysis for hundreds of haloes in the simulation and saved the neutrino overdensity maximum value together with its location. These values are shown in Figure~\ref{fig:pos_centre_array} as a function of the halo mass. The mean value of each parameter and the error bars corresponding to a $1\sigma$ error are represented. One could say that the effect is observable for halo masses greater than $3\times10^{14}\,h^{-1}M_\odot$. For those haloes, the mean position of the neutrino overdensity maximum is at $(-0.06,-0.06)\ h^{-1}$Mpc. 

\section{Discussion and conclusions}
\label{sec:discussion}
High-resolution cosmological simulations open up the possibility to analyze in detail specific high precision features of the cosmological model. In this work we have focussed on the properties of neutrino haloes, looking for asymmetries in regions away from the centre of the haloes but also in regions close to it. To this end, we have used the HR-DEMNUni simulations. These kinds of studies are motivated by the current context of new experiments aimed at improving our understanding of the Universe. If work continues in this direction, it is possible that, in the future, neutrinos from the cosmic background will be detected. Some evidences of the effects studied in the present work will help to make sure that neutrinos are being detected. The simulations used in this work allow us to obtain more reliable results as they contain neutrinos with a total mass closer to the actual limits established by cosmology. 

We have obtained angle-averaged neutrino profiles for several halo masses, using a kernel to obtain a smoother shape, similar to previous works, but this time for a lower neutrino mass. We have shown that, due to the higher thermal velocity of lower mass neutrinos, the clustering in haloes of even $M\sim10^{14}$\hmsun is more suppressed. The lower the neutrino mass the more diffuse the neutrino halo core is, then problems arise when fitting to the formula proposed. We put the limit of the applicability of this formula in halos of $4 \times10^{14}$\hmsun. This result suggest that low mass neutrinos can be understood mostly as light-like particles in the surrounding of dark matter haloes with masses less than $4 \times10^{14}$\hmsun.
With lower neutrino mass simulations that fit the current cosmological limits close to the mass split from underground laboratories of $0.06$ eV, this limit could be pushed further.

We have also investigated asymmetries that can arise when the neutrino profiles are not averaged over the solid angle. As we know, cosmic neutrinos will cluster around virialized dark matter haloes as a consequence of the huge gravitational field produced by clusters of $10^{11} - 10^{15}\,h^{-1}M_\odot$ of mass. As neutrinos will fall-in following the gravitational field, if this field is not isotropic enough we could find some anisotropies in neutrino field as well. In that sense, in the region located ahead of the dark matter halo centre we expected a neutrino density higher than in  the region located behind, with respect to the direction of CDM motion, i.e., along the anisotropy. In this work, we have shown that the predicted front loading of neutrinos effect is indeed present in the simulations and has been quantified. It is visible for $\text{R}>6\,h^{-1}\rm{Mpc} $ for a solid angle $\Delta\theta=60^\circ$, $\Delta\phi=120^\circ$ centred at the centre of the halo. We have also quantified its mass dependence, the result obtained is an average overdensity around 0.1 and independent of the halo mass. Only for haloes with masses close to $10^{15}\,h^{-1}M_\odot$ some enhancement of the effect is observed.

We found that the HR-DEMNuni simulations have enough resolution to search for the wakes in neutrino profiles recently proposed in~\cite{Zhu,LoVerde}. These wakes were expected in regions close to the centre of the haloes and due to the relative motions of neutrinos and CDM. We searched for them in the innermost regions within the haloes, in a box of $6\ h^{-1}$Mpc side placed in the centre of the halo, and we found that these wakes only take place in the most massive haloes ($M_{\rm h} > 3\times10^{14}\, h^{-1}M_\odot$). An average displacement of the neutrino overdensity distribution maximum of $0.06\,$\hmpc was obtained.

An experimental detection of all these effects will cement the case for neutrinos being observed in the sky and in addition, detailed study of these effects will in turn unveil the nature of neutrinos and the Universe when it was only one second old. Finally, we conclude that these very fascinating features will be seen once cosmic neutrino background experiments manage to distinguish direction and angle.

\acknowledgments 
The DEMNUni simulations were carried out in the framework of ``The Dark Energy and Massive Neutrino Universe" project, using the Tier-0 IBM BG/Q Fermi machine, the Tier-0 Intel OmniPath Cluster Marconi-A1 and Marconi-100 of the Centro Interuniversitario del Nord-Est per il Calcolo Elettronico (CINECA). CC acknowledges a generous CPU and storage allocation by the Italian Super-Computing Resource Allocation (ISCRA) as well as from the coordination of the ``Accordo Quadro MoU per lo svolgimento di attività congiunta di ricerca Nuove frontiere in Astrofisica: HPC e Data Exploration di nuova generazione'', together with storage from INFN-CNAF and INAF-IA2. This work was supported by the “Center of Excellence Maria de Maeztu 2020-2023” award to the ICCUB (CEX2019-000918-M) funded by MCIN/AEI/10.13039/501100011033. Funding for the work of RJ was partially provided by project PID2022-141125NB-I00.


\providecommand{\href}[2]{#2}\begingroup\raggedright\endgroup

\end{document}